\documentclass[twocolumn]{raa}
\usepackage{caption,subcaption}
\usepackage{graphicx,times}          
\usepackage{natbib}
\usepackage{overpic}
\usepackage{ulem} 
\graphicspath{{./}{figures/}}
\usepackage{amssymb,amsmath}
\usepackage{float}
\usepackage{multirow}

\bibpunct{(}{)}{;}{a}{}{,}
\usepackage{hyperref}
\hypersetup{colorlinks,				
	linkcolor=blue,
	citecolor=blue}
	
\def\be{\begin{equation}}
\def\ee{\end{equation}}
\def\bea{\begin{eqnarray}}
\def\eea{\end{eqnarray}}

\def\ciii{C\,{\sc iii]}}
\def\civ{C\,{\sc iv}}
\def\nv{N\,{\sc v}}
\def\mgii{Mg\,{\sc ii}}

\begin{document}

\title{
Variations of Broad Emission Lines from periodicity QSOs under the interpretation of supermassive binary black holes with misaligned circumbinary broad line regions}

\author{Xiang Ji
\inst{1,2}
\and Junqiang Ge
\inst{1}
\and Youjun Lu
\inst{1,2}
\and Changshuo Yan
\inst{1,2}
}
\institute{CAS Key Laboratory for Computational Astrophysics, National Astronomical Observatories, Chinese Academy of Sciences, 20A Datun Road, Beijing 100101, China \email{jqge@nao.cas.cn (Junqiang Ge); luyj@nao.cas.cn (Youjun Lu)}
School of Astronomy and Space Science, University of Chinese Academy of Sciences, No. 19A Yuquan Road, Beijing, 100049, China
}

\abstract{
Quasars with periodic light curves are considered as candidates of supermassive binary black hole (BBH) systems. One way for further confirmations may be searching for other characteristic signatures, such as those in their broad emission lines (BELs), if any, which require a thorough understanding on the response of BELs to the BBH systems. In \cite{Ji2021}, we have investigated the response of circumbinary broad line region (BLR) to the central active secondary black hole under the relativistic Doppler boosting (BBH-DB) and intrinsic variation (BBH-IntDB) dominant mechanisms for continuum variation by assuming the middle plane of the BLR aligned with the BBH orbital plane. In this paper, we explore how the BEL profiles vary when the BLR is misaligned from the BBH orbital plane with different offset angles under both the BBH-DB and BBH-IntDB scenarios. Given a fixed inclination angle of the BBH orbital plane viewed in edge-on and similar continuum light curves produced by the two scenarios, increasing offset angles make the initial opening angle of the circumbinary BLR enlarged due to orbital precession caused by the BBH system, especially for clouds in the inner region, which result in Lorentz-like BEL profiles for the BBH-DB model but still Gaussian-like profiles for the BBH-IntDB model at the vertical BLR case. The amplitude of profile variations decrease with increasing offset angles for the BBH-DB scenario, while keep nearly constant for the BBH-IntDB scenario, since the Doppler boosting effect is motion direction preferred but the intrinsic variation is radiated isotropically. If the circumbinary BLR is composed of a coplanar and a vertical components with their number of clouds following the mass ratio of the BBHs, then the bi-BLR features are more significant for the BBH-IntDB model that require larger mass ratio to generate similar continuum variation than the BBH-DB model. 
\keywords{Black hole physics; Quasars; Supermassive black holes; emission lines; profiles}
}

\authorrunning{Ji et al.}

\titlerunning{Variations of BEL profiles in misaligned BBH and BLR systems}

\maketitle

\section{Introduction}

Searching for supermassive binary black holes (BBHs) at different separations are vital for understanding the hierarchical merging processes of galaxies, the formation and evolution of BBHs  \citep[e.g.,][]{1980Natur.287..307B, Yu02, 2003ApJ...582..559V}, and their gravitational wave radiation \citep[e.g.,][]{2009MNRAS.394.2255S, CYL20}. Merging galaxies with dual active nucleus separated at kiloparsec scale have been intensively studied both in observations \citep[e.g.,][]{ 2009ApJ...702L..82C, 2013ApJ...777...64C, 2015ApJ...806..219C, 2011ApJ...737..101L, 2018ApJ...862...29L,  2011ApJ...735...48S, 2012ApJS..201...31G, 2012ApJ...746L..22K, 2021A&A...646A.153S} and theory \citep[][]{2005Natur.433..604D, 2011ApJ...738...92Y,2012ApJ...748L...7V,2013MNRAS.429.2594B,2015MNRAS.447.2123C,2017MNRAS.469.4437C,2016MNRAS.458.1013S,2019RAA....19..177Y, 2019SCPMA..6229511Y}. At parsec scale, one BBH was identified by radio observations \citep[e.g.,][]{2006ApJ...646...49R}, though such cases may be rare. Subparsec BBHs are currently difficult to spatially resolve because of the angular resolution limitation of available facilities \citep[e.g.,][]{Yu02, 2011MNRAS.410.2113B}, but may be revealed by a number of proposed spectral signatures \citep[see an overview by][]{2020RAA....20..160W}. These signatures include periodically varying double-peaked or asymmetric broad emission lines (BELs) \citep[e.g.,][]{BL09, Tsalmantza11, 2012ApJ...759..118B, Eracleous12, 2010ApJ...725..249S,      2012NewAR..56...74P,  2016ApJ...822....4L} and UV/optical light curves \citep[e.g.,][]{1988ApJ...325..628S, Graham15, 2015MNRAS.453.1562G, 2016MNRAS.463.2145C, 2018MNRAS.476.4617C, 2016ApJ...822....4L, 2019ApJS..241...33L},  optical-UV continuum deficiency  \citep[e.g.,][]{2015ApJ...809..117Y, Zheng16}, and changing-look AGNs \citep[e.g.,][]{2020A&A...643L...9W}. However, it is still hard to confirm the BBH candidates suggested by these signature, as their alternative interpretations are hard to be ruled out \cite[e.g.,][]{2004ApJ...613L..33G, 2012NewAR..56...74P}.

The periodicity QSOs (including the archetype PG 1302-102; \citealt{Graham15, 2015MNRAS.453.1562G, 2016MNRAS.463.2145C, 2018A&A...615A.123Q, 2020MNRAS.499.2245C, 2020MNRAS.496.1683X, 2021MNRAS.500.4025L}) are an intriguing subset of the currently known BBH candidates. In the BBH scenario for these QSOs, the periodicity may be either due to the Doppler boosting (DB) of the continuum emission from  accretion onto the secondary component modulated by its orbital motion \citep[][]{2015Natur.525..351D, 2020ApJ...901...25D} or intrinsic variation of the accretion rate modulated by BBH orbital motion \citep[e.g.,][]{2008ApJ...682.1134H, 2008ApJ...672...83M,  2014MNRAS.439.3476R, 2015MNRAS.447L..80F,  2017ApJ...838...42B, 2018ApJ...853L..17B}. However, the periodical variation itself may be not the definitive evidence for the existence of BBHs in periodicity QSOs. Furthermore, the periodicity of PG1302-102 was even doubted by subsequent observations \citep{2018ApJ...859L..12L}.

Among a large sample of QSOs, long-term stochastic variability may also result in a short-time periodicity in a small fraction of the sample \citep[][]{2016MNRAS.461.3145V}, similar as that found in \citet{2015MNRAS.453.1562G} and \citet{2016MNRAS.463.2145C}. In order to identify or rule out these BBH candidates, therefore, it may be necessary to find other BBH signatures for these QSOs and investigate whether their spectral properties are consistent with the BBH interpretation. 

\cite{Song2020} analyzed the BEL properties of periodicity QSOs in the sample of \cite{2016MNRAS.463.2145C} and found that the BELs of all these QSOs are circum-binary and viewed at an inclination angle of $i_{\rm BLR}<45^\circ$ while at least half of these BBH candidates are viewed at an orientation close to edge-on if its periodicity is due to the DB effect. These apparent contradictory results suggest that the DB effect is not the main reason for the periodicity or the BLR is mis-aligned with BBH orbital plane in those BBH systems viewed at an orientation close to edge-on \citep[e.g.,][]{Song2021}. It is therefore of great importance to study the behaviour of BELs for the periodicity QSOs under the BBH interpretation.

\citet[][hereafter PAPER I]{Ji2021} performed a systematic analysis on the BEL profile variations of BBH systems with circumbinary BLRs, focusing on the cases of BLR aligned with BBH orbital plane. They investigated the responses of BELs to the continuum variation under two scenarios: 1) BBH-DB, i.e., the flux variation is solely due to the DB effect, and 2) BBH-IntDB, i.e., the flux variation is mainly due to the intrinsic variation but with some contribution from the DB effect. They found that the variation pattern of BELs resulting from these two scenarios are significantly different from each other, since the DB effect has a preferred direction along the motion direction of the secondary BH, while the intrinsic variation dominant case does not. The differences in the periodic profile variations may offer a robust way for identifying/falsifying the BBH candidates and distinguish different mechanisms for the periodicity.

In this work, we extend the studies done in PAPER I to consider the BBH systems with BLRs misaligned from the BBH orbital plane  and investigate whether the properties of BELs from such BLR-BBH systems and its variation are different from the case of BLRs aligned with BBH orbital plane and can be used to distinguish different periodicity mechanisms. 
The paper is organized as follows. In Section 2, we introduce our model construction and parameter settings of the misaligned BLR and BBH systems. In Section 3, we analyze the response of BLR emissions and BEL profile variations for these offset systems. Discussions are made in Section 4 and conclusions are summarized in Section 5.

\section{Simple models for misaligned BBH and BLR system}

The misaligned BBH and BLR systems are built by following the method introduced in PAPER I, in which we have compared the profile differences caused by the BBH-DB, BBH-IntDB, and BH-Int scenarios, where the BBH-DB scenario corresponds to the relativistic Doppler boosting/weakening effects from a BBH system \citep[e.g.,][]{2015Natur.525..351D}, and the BBH-IntDB scenario is for the continuum variation mainly caused by accretion rates modulated by the BBH motion \citep[e.g.,][]{2008ApJ...682.1134H, 2008ApJ...672...83M,  2014MNRAS.439.3476R, 2015MNRAS.447L..80F,  2017ApJ...838...42B, 2018ApJ...853L..17B} with some contribution from the DB effect, and the BH-Int scenario is the intrinsic continuum variation in single BH system set for comparison. Instead of the co-planar BBHs and BLRs studied in PAPER I, here we investigate how the BELs from offset BLRs response to the continuum variation under the BBH-DB or BBH-IntDB scenarios. 

\subsection{Configuration of misaligned BBH and circumbinary BLR systems}

\begin{figure}[ht]
\centering
\includegraphics[width=0.49\textwidth]{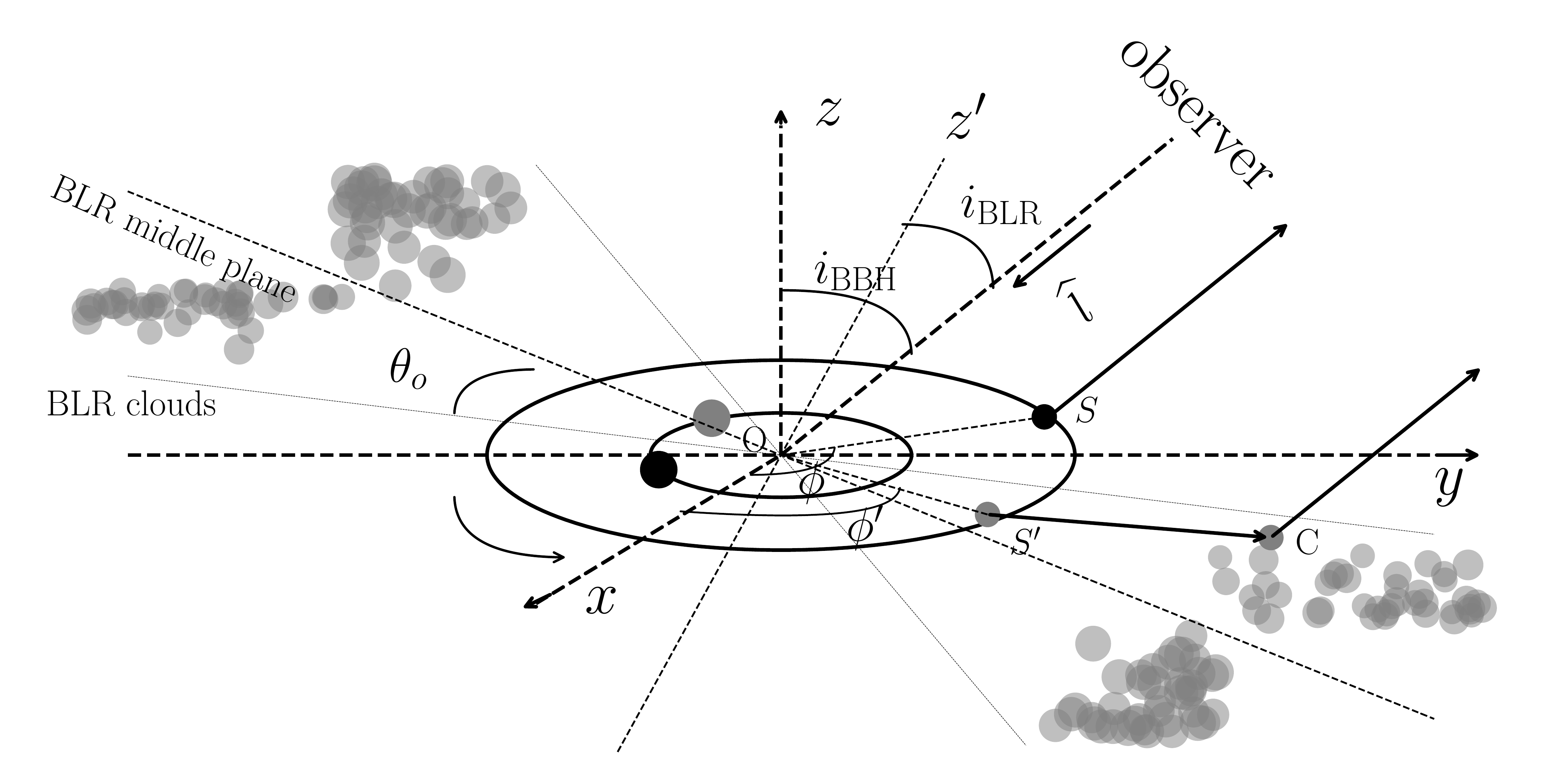}
\caption{Sketch diagram for the geometry of a BBH system with an offset circumbinary BLR. The primary (large solid circles) and secondary (small solid circles) BHs are supposed to counter-clockwise rotate on circular orbits in the $xy$-plane around the mass center, $i_{\rm BBH}$ denotes the inclination angle of the BBH orbital plane, i.e., the angle between the direction of the line of sight (LOS) and the $\vec{z}$-axis. The black and grey solid circles represent the positions of the two components at two observational time. Grey points around the offset BLR middle plane represent BLR clouds, rotating around the BBH system within an opening angle of $\theta_{\rm o}$. C represents an arbitrary cloud in the BLR, $i_{\rm BLR}$ is the inclination angle of the BLR middle plane, which is the angle between the direction of the LOS and the $\vec{z'}$-axis, namely, the normal direction of the BLR middle plane. At a specific observation time $t_{\rm obs}$, the observer receives continuum photons emitted from the disk around the secondary BH at $S$ and the orbital phase is $\phi$. At the same time, the observer also receives BEL photons re-emitted from the cloud C, which were induced by the illumination of continuum flux emitted from the disk around the secondary BH at $S^{\prime}$ with orbital phase of $\phi'$.
}
\label{fig:f1}
\end{figure} 

Figure~\ref{fig:f1} shows the sketch diagram of a misaligned circumbinary BLR and BBH system, which is similar to the Figure 1 in PAPER I except for changing the offset angle $\Delta i$ from co-planar to misaligned, where $\Delta i$ is defined as the angle between the normal vectors of the BLR middle plane and BBH orbital plane. For the circumbinary offset BLR, the radial emissivity distribution of BLR clouds follows a shifted $\Gamma-$distribution as proposed by \cite{2014MNRAS.445.3055P}, which is consistent with 3C 273 observed by VLT/GRAVITY \cite{2018Natur.563..657G}:
\begin{equation}
R_{\rm ga} = R_{\rm S} + F R_{\rm BLR} +g(1-F) \beta^{2} R_{\rm BLR},
\end{equation}
where $R_{\rm S} $ is the Schwarzschild radius, $R_{\rm BLR}$ is the mean BLR radius, F = $R_{\rm min}/R_{\rm BLR}$ is the fractional inner BLR radius, $\beta$ is the shape parameter, and $g = p(x|1/\beta^{2},1)$ is drawn randomly from a Gamma distribution. Here we simply assume the $R_{\rm BLR}$ of BBH systems follow the same empirical relation $R_{\rm BLR} \propto L^{1/2}$ as that for single BH systems \citep[][]{2000ApJ...533..631K, 2007ApJ...659..997K, 2002MNRAS.337..109M, 2004ApJ...613..682P}:
\begin{equation}
R_{\rm BLR}  \approx 2.2 \times 10^{-2}   \left (\frac{\lambda_{\rm Edd}}{0.1} 
\right) ^{1/2} \left (\frac{M_{\bullet}}{10^{8} M_{\odot}} \right)^{1/2} {\rm pc}.
\label{eq:RBLR}
\end{equation}
where $\lambda_{\rm Edd}$ is the Eddington ratio, $M_{\bullet}$ is the mass of the active secondary BH.

By applying the calculation method of restricted three body dynamics to the BBHs and BLR cloud system as described in PAPER I, we record the positions and velocities of BLR clouds with bound elliptical orbits in a dynamic stabilization state (see detailed description in PAPER I). One important feature that is significantly different from the coplanar case assumed in PAPER I is that, the orbital precession for BLR clouds closer to the BBH system would be more significant, which can enlarge the opening angle of the BLR especially for inner BLR regions due to the orbital precessions caused by the corotating BBHs.

The response of a BLR cloud $C_i$ emission to the central periodical continuum is the same as that introduced in Section 2.1 of our PAPER I, with the central radiating source has periodical variation induced by the DB enhancement/weakening in the BBH-DB case, and periodical intrinsic accretion rate variation dominated with the DB effect as secondary effect to the variability in the BBH-IntDB scenario (see Section 2.2 of PAPER I). Here we also use a power-law $F_{\nu, \rm e}(\nu; t) \propto \nu^{\alpha}$ to describe the intrinsic continuum flux emitted from the disk around the secondary BH. The amplitude of the periodical variation caused by the Doppler boosting ($\mathcal{A}_{\rm DB}$) can be given by
\begin{eqnarray}
\log \mathcal{A}_{\rm DB} & \sim & \frac{1}{2} \left[ \log \left| D^{3-\alpha}_{\rm sec,max}-1\right| + \log \left| D^{3-\alpha}_{\rm sec,min} -1 \right|\right], \nonumber \\ 
\label{eq: DBamp}
\end{eqnarray}
where $D_{\rm sec,max}$ and $D_{\rm sec,min}$ are the maximum and minimum Doppler boosting factor at $\phi=0^\circ$ and $180^\circ$, respectively.
Given the amplitude of the observed continuum variation $\mathcal{A}$, for the BBH-IntDB scenario, the contribution fraction of the intrinsic variation becomes
\begin{equation}
\mathcal{A}_{\rm Int} = \frac{\mathcal{A}+1}{\mathcal{A}_{\rm DB}+1} -1.
\label{eq:Intamp}
\end{equation}

For the emission mechanisms of BLR clouds, we take into consider 1) the time-delayed response caused by the continuum radiated in the BBH-DB and BBH-IntDB scenarios (see Section 2.4 of PAPER I); and 2) the gravitational redshift of photons emitted from BLR clouds (see Section 2.5 of PAPER I). The radiation of BLR clouds can be described as (see Equation (21) in Section 2.5 of PAPER I for more details):
\begin{eqnarray}
L(v,t_{\rm obs})
& \propto & \left. \sum_{i=1}^{N_{\rm tot}} F_{\nu,\rm e}(\nu; t_{\rm in}^{'}) D^{3-\alpha}_{{\rm C}_i2}\frac{|\vec{r}_{{\rm C}_i}|^{2}}{|\vec{r}_{{\rm C}_i} - \vec{r}_{\rm sec}|^2}  \right|_{v= { v^{\rm tot}_{{\rm C}_i}}   }. 
\label{eq:BBHLp}
\end{eqnarray}
where $N_{\rm tot}$ is the total number of BLR clouds, $D_{{\rm C}_i2}^{3-\alpha}$ represents the enhancement/weakening of the ionizing flux $F_{\nu,\rm e}(\nu; t_{\rm in}^{'})$ emitted at $t_{\rm in}^{'}$ and received by the BLR cloud at $t_{\rm obs}$ due to the relativistic motion of the secondary BH, and the term $|\vec{r}_{{\rm C}_i}|^2/|\vec{r}_{{\rm C}_i} -\vec{r}_{\rm sec}(t'_{\rm in}) |^2$ considers the variation of the ionizing flux due to the  position change of the source (secondary BH), $v^{\rm tot}_{{\rm C}_i} \simeq v^{\rm D}_{{\rm C}_i} + v^{\rm g}_{{\rm C}_i} $ represents the summation of the Doppler redshift/blueshift  $v^{\rm D}_{{\rm C}_i}=\left\{\left[ \gamma_{{\rm C}_i{\rm obs}}(1-\vec{\beta}_{{\rm C}_i{\rm obs}}\cdot \boldsymbol{\hat{l}})\right]^{-1}-1\right\} c$  and the gravitational redshift of photons emitted from an individual BLR cloud received by the distant observer $v^{\rm g}_{{\rm C}_i} =  GM/(r_{\rm C_{i}}c) $  \citep[][]{2014ApJ...794...49T}, $\vec{\beta}_{{\rm C}_i{\rm obs}}=\vec{v}_{{\rm C}_i{\rm obs}}/c$ is the observed velocity of the cloud C$_i$ in unit of the light speed, $\gamma_{{\rm C}_i{\rm obs}}=1/\sqrt{1-|\vec{\beta}_{{\rm C}_i{\rm obs}}|^2}$, $M$ is the total mass of the central BBH system, and $\vec{r_{\rm C_{i}}}$ is the distance vector of the BLR cloud to the mass center of the BBH system. In this work, we also set the spectral index $\alpha=-2$ \citep[e.g.,][]{2015Natur.525..351D} as done in PAPER I for simplicity. In each model, the line emissivity is assumed to be proportional to the flux received by each cloud.

\subsection{Model settings}
\label{sec:modelset}
\begin{table*}
\centering
\caption{
Model Parameters for BBH system with a (circumbinary) BLR under different offset angles and optical/UV periodicity scenarios.
}
\begin{tabular}{lccccccccccc} \hline 
\multirow{2}{*}{Model} & $M_{\bullet\bullet}$ & Mass &\multirow{2}{*}{$T_{\rm orb}$} & $a_{\rm BBH}$ &\multirow{2}{*}{$i_{\rm BBH}(^\circ)$} & \multirow{2}{*}{$\lambda_{\rm Edd}$} &  \multirow{2}{*}{$\mathcal{A}$} & \multirow{2}{*}{$\mathcal{A}_{\rm Int}$} & \multicolumn{3}{c}{BLR}\\ \cline{2-2} \cline{5-5} \cline{10-12}
& $(10^9M_{\odot})$ & Ratio & &($10^{-3}$\,pc) & & & &   
& $R_{\rm BLR}$\,(pc)  & $\theta_{\rm o}^{\rm ini} (^\circ)$  & $\Delta i~(^\circ)$ \\ \hline \hline
BBH-DB & $5$   & 0.2     &2       & 13  &85   & 0.09    &0.57  & 0     & 0.06   & 30 & 0/30/60/90 \\
BBH-IntDB    & $0.5$ & 0.8     &2       & 6  &85    &0.9   &0.57 &0.34  & 0.06   & 30 & 0/30/60/90  \\  \hline \hline
\end{tabular}
\flushleft
Note: Columns from left to right list the model name, the total mass of the BBH system $M_{\bullet\bullet}$, the variation period of the continuum or the orbital period of the BBH system $T_{\rm orb}$, the BBH semimajor axis $a_{\rm BBH}$, the inclination angle of BBH orbital plane $i_{\rm BBH}$, the Eddington ratio of the BBH accretion system $\lambda_{\rm Edd}$, the total variation amplitude $\mathcal{A}$, the amplitude of the intrinsic variation $\mathcal{A}_{\rm Int}$, the mean BLR size $R_{\rm BLR}$, the initial opening angle of the BLR $\theta_{o}^{\rm ini}$, and the offset angle $\Delta i$ of the BLR middle plane relative to the BBH orbital plane. 
\flushleft
\label{tbl:t1}
\end{table*}

With the model of misaligned BBH and BLR system described above, we then set model parameters to explore features of BEL profiles in the misaligned case for both the BBH-DB and BBH-IntDB scenarios, which can guide us to distinguish them by spectroscopic observations. For simplicity, we assume the same model parameters as the BBH-DB-hi and BBH-IntDB-hi models listed in our PAPER I, but changing the offset angle of the BLR middle plane misaligned from the BBH orbital plane with $\Delta i = 0^\circ$ to $90^\circ$, which means that the BLR can be viewed from $i_{\rm BLR}\sim 0^\circ$ to $\sim 90^\circ$ and we hence define the two models as BBH-DB and BBH-IntDB to avoid misunderstanding. In Table \ref{tbl:t1}, we list all model parameters in columns from left to right, only with $\Delta i$ varying from $0^\circ$ to $90^\circ$ that is different from the co-planar case ($\Delta i=0^\circ$) assumed in PAPER I. All the BLR clouds are set to be co-rotating with the BBH system in the counter-clockwise direction. Due to the orbital precession of BLR clouds caused by the corotating BBH system, the initial opening angle of the BLR $\theta_{\rm o}^{\rm ini}$ will be enlarged with increasing offset angles, especially for those clouds nearby the BBH system. For the BBH-DB scenario, the opening angle of BLRs in the inner region ($r<R_{\rm BLR}$) at the dynamical stabilization state can increase from $\theta_{\rm o}=60^\circ$ at $\Delta i=30^\circ$ to almost spherical at $\Delta i = 90^\circ$, while $\theta_{\rm o}$ at the outer region ($r>R_{\rm BLR}$) is enlarged less than half of the magnitude compared to the inner region. For the BBH-IntDB scenario, since the $R_{\rm BLR}/a_{\rm BBH}$ is $\sim 2$ times larger than that of the BBH-DB model, the orbital precessions broaden the opening angle of the inner region $r<R_{\rm BLR}$ to the comparable amplitude as the outer region of the BBH-DB model, and the opening angle of the out region in the BBH-IntDB model are not broadened significantly. Therefore, the real opening angles of the circumbinary BLRs are not uniform and depend on both the offset angle and radii that the clouds located. 

The BBH-DB and BBH-IntDB models are all assumed to have the same BBH orbital period, i.e., $T_{\rm orb}=2$ yr, as the typical period found in \citet{2015MNRAS.453.1562G} and \citet{2016MNRAS.463.2145C}. We assume that the BBHs rotate around each other on circular orbit, with the semimajor axis $a_{\rm BBH}$ determined by the mass of the BBH system ($M_{\bullet\bullet}$), mass ratio, and $T_{\rm orb}$. 

To keep the observed continua of the two models in Table \ref{tbl:t1} have not only the same variation period but also the same amplitude, we then set $\lambda_{\rm Edd}=0.09$ for BBH-DB systems and $0.9$ for BBH-IntDB systems, which can derive the amplitude of the light curve $\mathcal{A}=0.57$. For the BBH-IntDB model, the intrinsic variation contributes a higher fraction of the total amplitude ($\mathcal{A}_{\rm Int}=0.34$) than the DB mechanism.

For the geometry of the misaligned BLR, we assume the mean radius follows the single BH case (Eq. \ref{eq:RBLR}) and derive $R_{\rm BLR}=0.06$ pc. Here we only consider flattened disk like BLR with an initial opening angle of $\theta_{\rm o}^{\rm ini}=30^\circ$, and assume the BLR can be misaligned from the BBH orbital plane with offset angles $\Delta i=0^\circ, 30^\circ, 60^\circ$, and $90^\circ$, which corresponding to the inclination angle of the BLR $i_{\rm BLR}=85^\circ, 65^\circ, 35^\circ$, and $5^\circ$. With the BBH-DB and BBH-IntDB models built, the line of sight (LOS) projected BEL profiles can hence be obtained.

\section{Results}
\label{sec:results}

The main goal of this paper is to investigate the BEL profiles and their variations for a circumbinary BLR misaligned from the BBH orbital plane with different offset angles, for both the BBH-DB and BBH-IntDB scenarios. The analyses start from interpreting the detailed response of BLR clouds to the central source, then go to the profile variations that can compare with spectroscopic observations directly. 

\subsection{Response of offset circumbinary BLRs to the continuum variation}

\begin{figure*}
\centering
\includegraphics[width=0.85\textwidth]{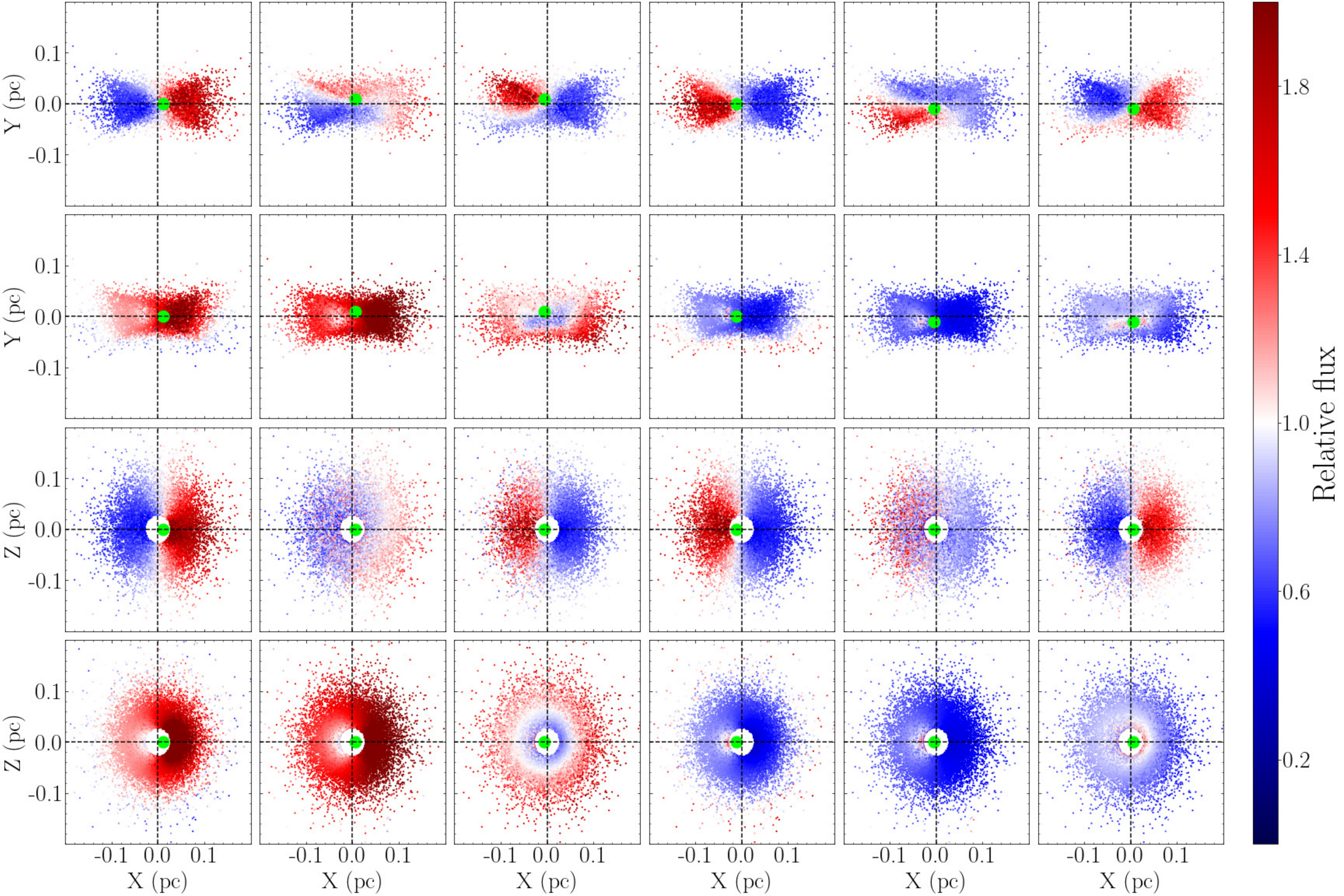}
\caption{Enhancement/weakening of ionizing flux received by individual BLR clouds, which are projected on the $xy$-plane (top two panels) and the $xz$-plane (bottom two panels) at six phases of the periodic optical/UV continuum variations for the BBH-DB (first and third rows) and the BBH-Int (second and fourth rows) scenarios at offset angle $\Delta i = 90^{\circ}$. Here we only consider the effect of DB and position variation for the BBH-DB case, and the intrinsic variation and position changes for the BBH-Int case, by setting the parameters of the BBH-Int model the same as that of the BBH-DB model. Panels from left to right describe the response of BLR clouds to the continuum variation at phases of $0$, $\pi/3$, $2\pi/3$, $\pi$, $4\pi/3$, and $5\pi/3$, respectively. The right colorbar marks the relative flux received by each cloud (represented by each point) in the BLR in unit of the mean flux at each giving observational moment. 
}
\label{fig:f2}
\end{figure*} 

The broad line emission from BLR clouds is affected by different mechanisms, i.e., the DB effect, position variation, and gravitational redshift for the BBH-DB scenario, the primary intrinsic variation with the secondary DB effect, position variation, and gravitational redshift for the BBH-IntDB scenario. Figure \ref{fig:f2} is plotted to clarify the different responses of BLR clouds to the DB effect and intrinsic variation. To compare the two effects directly, we take the parameters of the BBH-DB model listed in Table \ref{tbl:t1} to construct two models: 1) BBH-DB model by only including the DB effect and position variation, and 2) BBH-Int model with the same parameters as the BBH-DB case but only including the effect of intrinsic variation and position changes. The enhancement/weakening of ionizing flux from phases $0$ to $5\pi/3$ (left to right columns) present different response to the central source for the BBH-DB case presented in the first ($xy$-plane) and third rows ($xz$-plane), and for the BBH-Int case shown in the second ($xy$-plane) and fourth rows ($xz$-plane).

For the BBH-DB scenario, the strongest enhancement of the BLR emission appear in the moving direction of the secondary BH, BLR clouds with larger angular distance from the moving direction of the secondary BH are less affected by the DB effect.

As shown in the first and third rows of Figure \ref{fig:f2}, when the secondary BH rotating to the positive $x$-axis (phase 0, left column), the emissivity of (LOS projected) blue-shifted clouds are enhanced, and that of those red-shifted clouds are weakened. Once the secondary BH rotated to the negative $x$-axis (phase $\pi$, the 4th column), the ionizing flux of red-shifted BLR clouds are enhanced and that of blue-shifted ones are weakened. Since the travel time cross the BLR (0.06 pc) of the BBH-DB and BBH-Int models is substantially smaller than a quarter of the BBH orbital period (2 yr), BLR clouds in the blue- and red-shifted regions are hence enhanced or weakened alternatively due to the periodic modulation of the secondary BH rotation. For current BBH-DB case, the regions of BLR clouds enhanced/weakened by the DB effect and position variation are partly overlapped. 
For the BBH-Int case, the line emissivity enhancement/weakening propagate outward periodically due to the time-delay effect, but still with the feature of alternative enhancement/weakening for the blue- and red-shifted BLR clouds caused by the position variation, which becomes more significant with decreasing $R_{\rm BLR}/a_{\rm BBH}$. As presented in the second and fourth rows of Figure \ref{fig:f2}, the enhanced/weakened BLR clouds vary systematically along their radii, with cloud emission in the nearby regions of the secondary BH further enhanced compared to other ones.

\begin{figure*}
\centering
\includegraphics[width=0.85\textwidth]{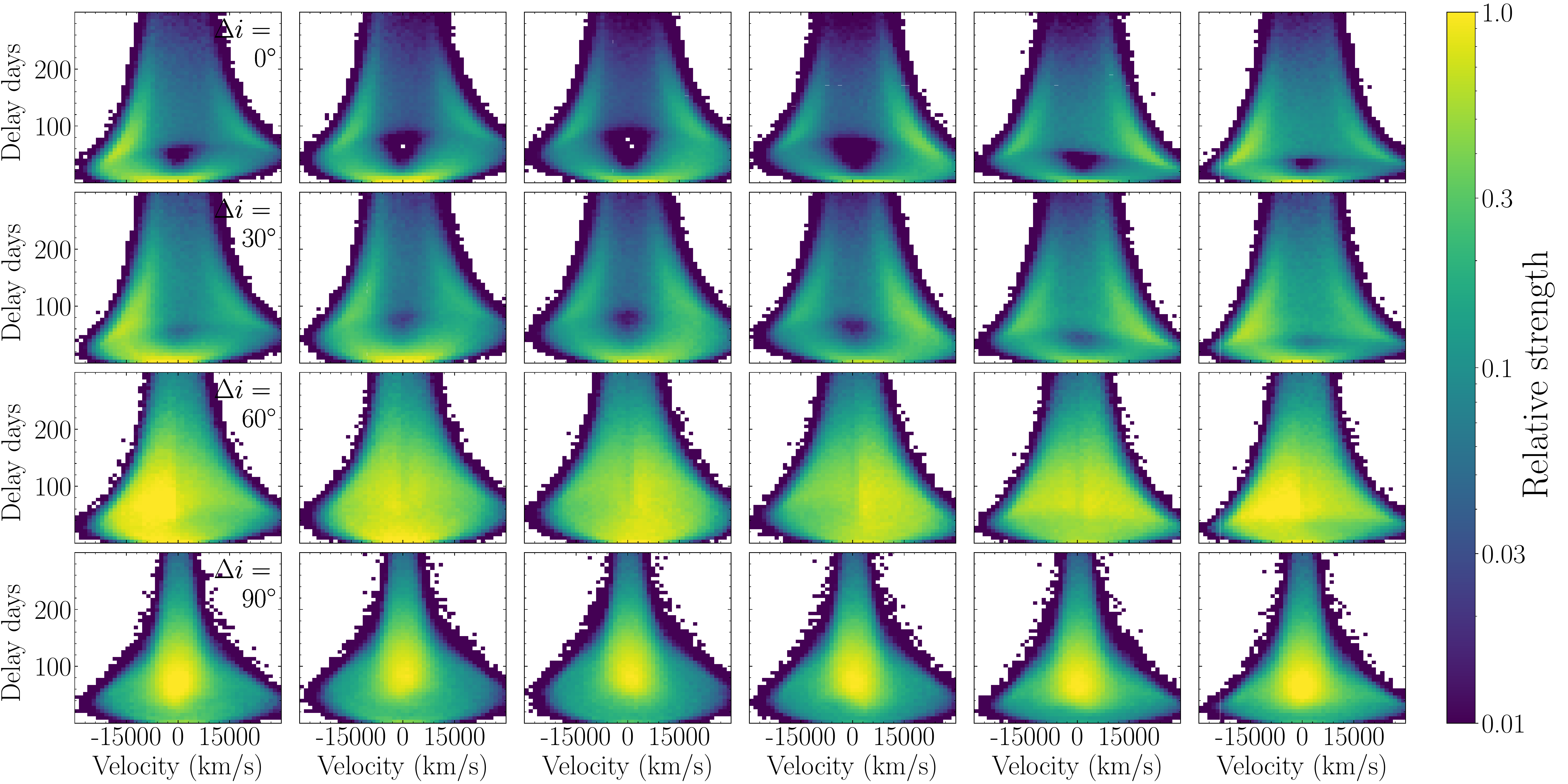}
\caption{Time delay versus LOS projected velocity for the BBH-DB scenario by including the DB effect, position variation, and gravitational redshift over a period of 2 yr. Top to bottom rows correspond to the offset angles increasing from $0^{\circ}$, 
$30^{\circ}$, $60^{\circ}$, to $90^{\circ}$. In each row, panels from left to right show the results obtained at six different orbital phases of the secondary BH, i.e., $0$, $\pi/3$, $2 \pi/3$,  $\pi$, $4 \pi/3$, and $5\pi/3$, respectively.  In each panel, the brightness of each pixel shows the relative value compared to the brightest one in the left panel of each separate row, with the detailed values labelled in the right colorbar.  
}
\label{fig:f3}
\end{figure*}

\begin{figure*}
\centering
\includegraphics[width=0.85\textwidth]{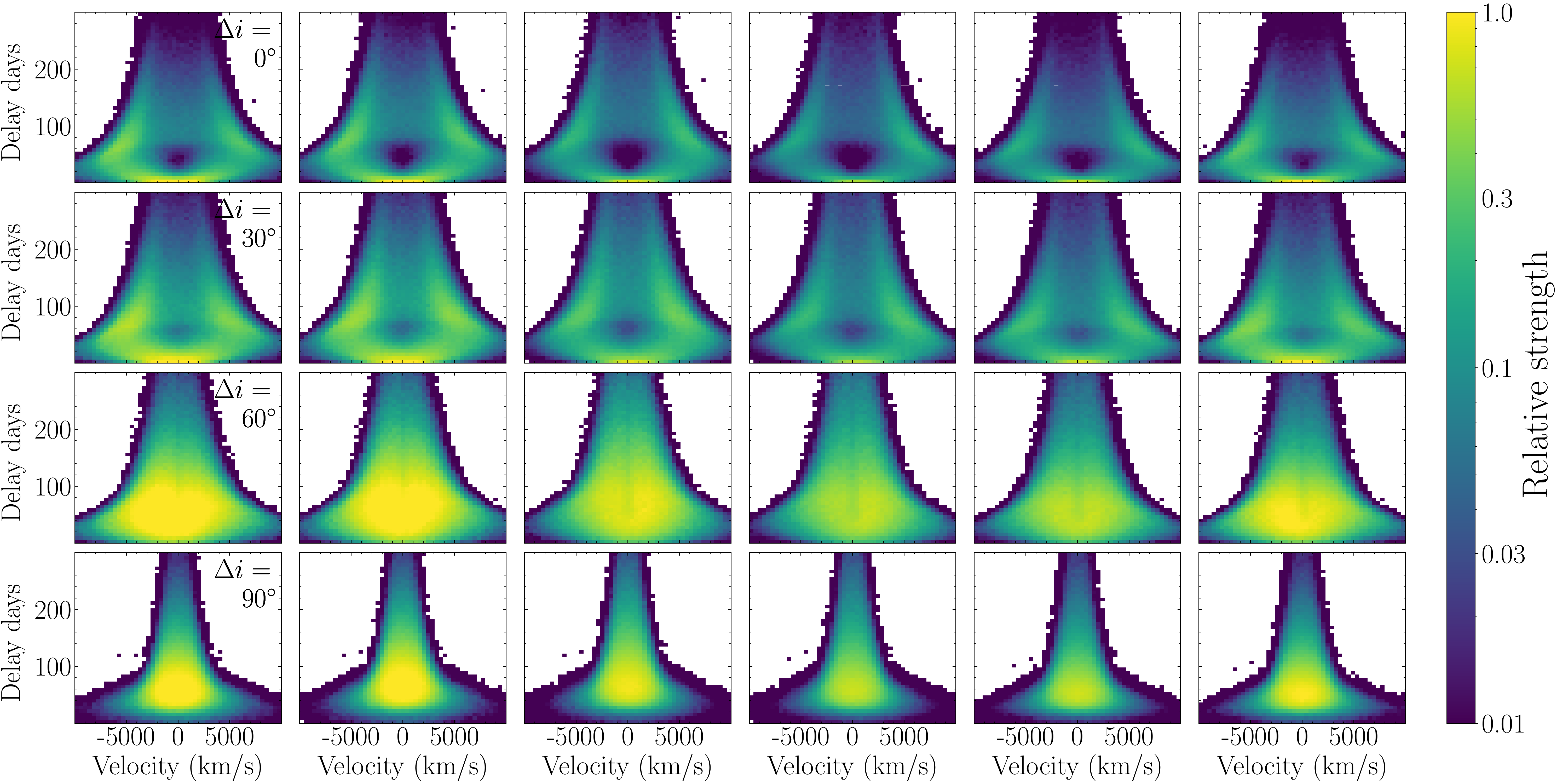}
\caption{Time delay versus LOS projected velocity maps derived by considering the effect of intrinsic accretion rate variation and position variation. Legends are similar to Fig.~\ref{fig:f3}, but rows from top to bottom are obtained with parameters of the BBH-IntDB model as listed in the second row in Tab.~\ref{tbl:t1}, also with $\Delta i$ varying from $0^{\circ}$, 
$30^{\circ}$, $60^{\circ}$, to $90^{\circ}$, respectively.
}
\label{fig:f4}
\end{figure*}

\begin{figure*}
\centering
\includegraphics[width=0.85\textwidth]{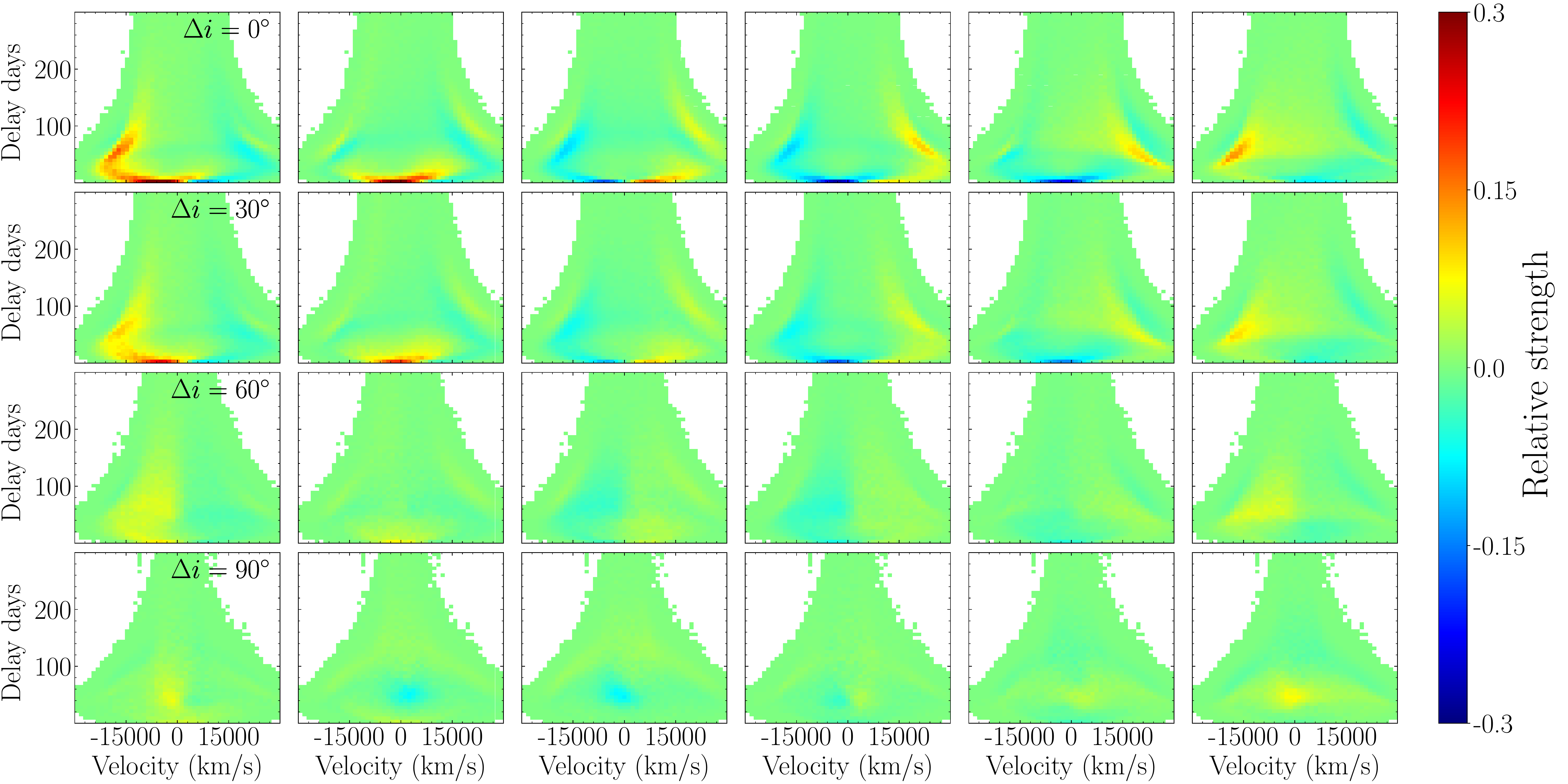}
\caption{
Differences of the time delay versus LOS projected velocity maps of Figure \ref{fig:f3}. Rows from top to bottom show the difference maps with offset angles $\Delta i=0^{\circ}$, $30^{\circ}$, $60^{\circ}$ and $90^{\circ}$, respectively. Panels from left to right shows the difference maps at phases $0$, $\pi/3$, $2\pi/3$, $\pi$, $4\pi/3$, and $5\pi/3$. The colorbar at the right side shows the relative flux variation in units of mean BEL flux in each separate row over a period of 2 yr by multiplying an arbitrary scaling factor for normalization.
}
\label{fig:f5}
\end{figure*} 
\begin{figure*}
\centering
\includegraphics[width=0.85\textwidth]{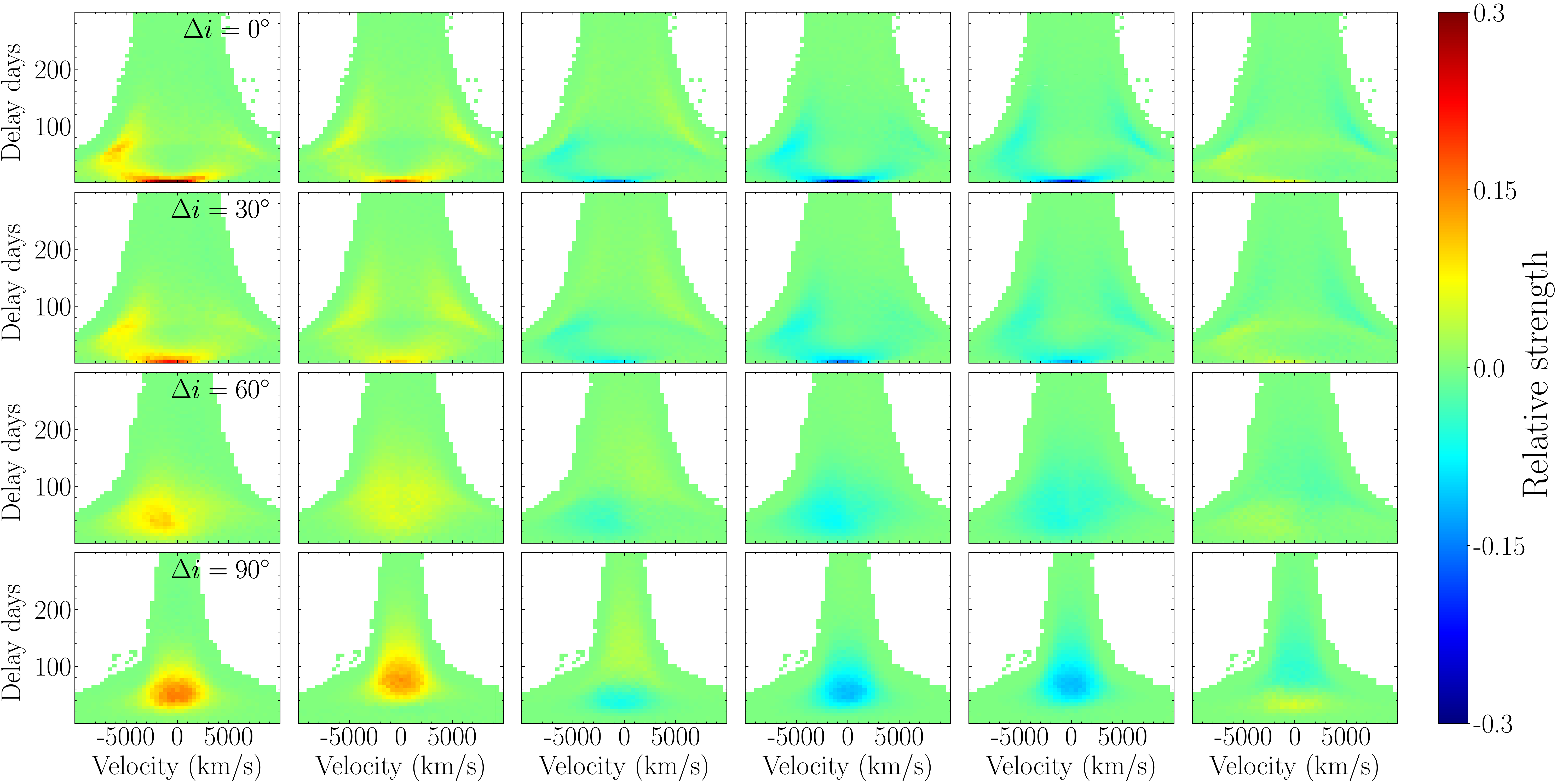}
\caption{Differences of the time delay versus LOS projected velocity maps of Figure \ref{fig:f4}. Legends are similar to Figure~\ref{fig:f5}. 
}
\label{fig:f6}
\end{figure*} 

\begin{figure*}[ht]
\centering
\includegraphics[width=0.98\textwidth]{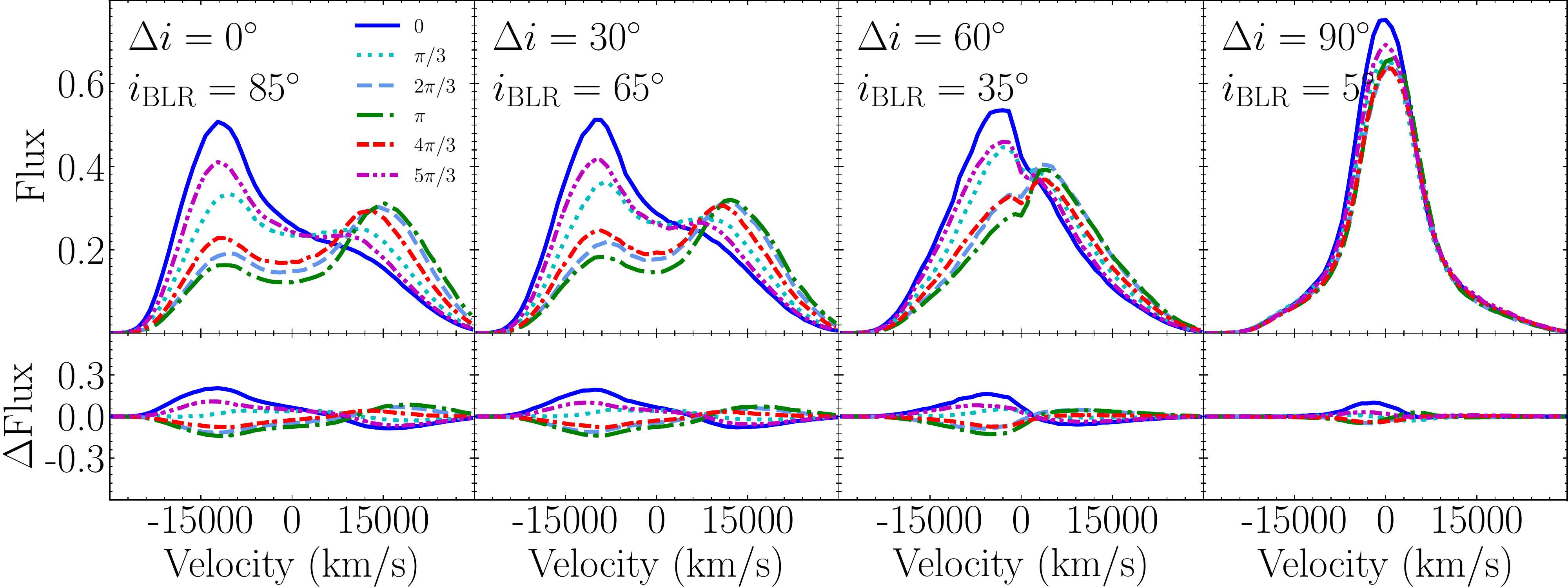}
\caption{Variations of BEL profiles over a single period (2 yr) of continuum variation for the BBH-DB model with $\theta_{\rm o}=30^\circ$, resulting from $\Delta i =0^{\circ} $ ($i_{\rm BLR}=85^\circ$, left column), $\Delta i=30^{\circ}$  ($i_{\rm BLR}=65^\circ$, middle-left column), $\Delta i=60^{\circ}$ ($i_{\rm BLR}=35^\circ$, middle-right column), and $\Delta i=90^{\circ}$ ($i_{\rm BLR}=5^\circ$, right column). The line flux is set in an arbitrary unit. In the top four panels, we show the profile variation in six different phases, i.e., $0$ (blue solid line), $\pi/3$ (cyan dotted line), $2\pi/3$ (light blue dashed line), $\pi$ (green dash-dotted line), $4\pi/3$ (red dash-dash-dotted line), and $5\pi/3$ (magenta dash-dot-dotted line), as labeled in the top-left panel. The bottom four panels show the corresponding difference of the BEL profiles at six phases from the mean line profile.}
\label{fig:f7}
\end{figure*} 

\begin{figure*}[ht]
\centering
\includegraphics[width=0.98\textwidth]{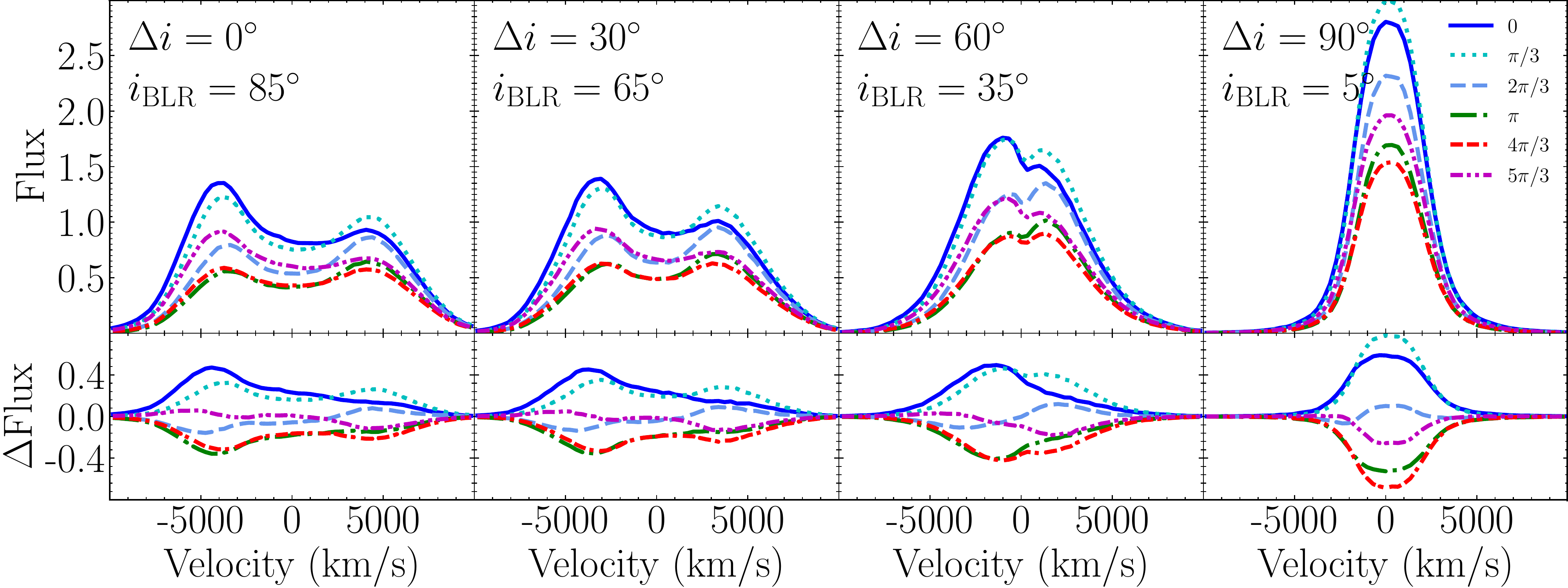}
\caption{
Variations of BEL profiles over a single period (2 yr) of continuum variation for the BBH-IntDB model with $\theta_{\rm o}=30^\circ$, with legend similar to Figure~\ref{fig:f7}.
}
\label{fig:f8}
\end{figure*}

After the detailed analyses on contributions of the DB effect and intrinsic variation to a misaligned BLR, Figures \ref{fig:f3} and \ref{fig:f4} show the two dimentional transfer functions (2DTFs) for the BBH-DB and BBH-IntDB models as listed in Table \ref{tbl:t1}. 

In Figure \ref{fig:f3}, to track the ionizing flux enhancement/weakening of BLR clouds and study how they rely on offset angles, we mark the relative brightness of each pixel compared to brightest pixel in the first panel of each separate row, and take four offset angles, i.e., $\Delta i=0^\circ, 30^\circ, 60^\circ$, and $90^\circ$, as shown from top to bottom rows for analysis.
By setting the observer to the $yz-$plane with $i_{\rm BBH}=85^\circ$ (Fig. \ref{fig:f1}), at each offset angle, the response of BLR clouds corotating with the BBH system is shown by six phases (left to right panels in Fig. \ref{fig:f3}) over a period. 

In the case of $i_{\rm BBH}=85^{\circ}$, with the increasing offset angle, the disk-like BLR with an initial opening angle of $\theta_{\rm o}^{\rm ini}=30^\circ$ is viewed closer to face-on, which cause smaller LOS projected velocities for the BLR clouds and also smaller relative velocities between the secondary BH and BLR clouds. On the other hand, with increasing $\Delta i$, the inclination and eccentricity of the orbit for a BLR cloud rotating around the BBH system will vary because of the orbital precession caused by the BBH system, which can increase the opening angle of the BLR and hence broaden the wings of observed BEL profile. Panels from top to bottom rows in Figure \ref{fig:f3} show decreasing LOS projected velocities with increasing $\Delta i$ (or decreasing $i_{\rm BLR}$) for BLR clouds with larger time delay. For those clouds closer to the BBH system with smaller time delay, their LOS projected velocities increase due to the increasing significance of orbital precessions, which is reflected especially by the third and fourth rows. For clearer comparison of the DB enhancement/weakening trends with increasing $\Delta i$, we plot the difference of BLR emissivity at the six phases in Figure \ref{fig:f5} for the BBH-DB model, from which we can see that larger offset angles correspond to lower amplitude of DB enhancement (in fading red) and weakening (in fading blue) than the case of $\Delta i=0$.

For the BBH-IntDB scenario, as shown in Figure \ref{fig:f4}, since the intrinsic accretion rate variation of the secondary BH takes the dominant effect and the DB effect only provides a secondary contribution, the emission pattern of BLR clouds mainly propagates outward periodically because of the time-delay effect, but still with the feature of alternative enhancement/weakening for the blue- and red-shifted BLR clouds caused by the DB effect. Figure \ref{fig:f6} clarifies this feature by showing the relative strength of BLR clouds in the BBH-IntDB model. At $\Delta i=0^\circ$, features are clear for both the periodic enhancement/weakening modulated by the intrinsic flux variation, with weak signatures that reflecting the periodic enhancement/weakening in the blue- and red-shifted regions modulated by the DB effect. With increasing $\Delta i$, the radially periodic enhancement/weakening due to the intrinsic variation and time-delay effect become more significant than the fading signature caused by the DB effect. At $\Delta i=90^\circ$ ($i_{\rm BLR}=5^\circ$), the feature of periodically enhanced/weakened BLR clouds in the blue- and red-shifted regions becomes insignificant. Since the mass of the BBH system in the BBH-IntDB model is 10 times smaller than that in the BBH-DB model, the corresponding $R_{\rm BLR}/a_{\rm BBH}$ is $\sim 2$ times larger than that of the BBH-DB model, which means that the orbital precession caused by the BBH system is systematically smaller than that in the BBH-DB scenario. This is directly reflected by comparing the cases of $\Delta i=60^{\circ}$ and $\Delta i=90^{\circ}$ shown in Figures \ref{fig:f3} and \ref{fig:f4}, and also Figures \ref{fig:f5} and \ref{fig:f6}, in which the BLR clouds at inner radii are redistributed.

\subsection{Dependence on different offset angles for the circumbinary BLR}

With the interpretation on detailed response of BLRs to the central varying continuum at different offset angles, we can explore detailed features of their periodic profile variations, which can be applied for fitting observed BELs in mostly UV, optical, and infrared bands, such as Ly$\alpha$, \nv, \ciii, \civ, \mgii, H$\beta$, H$\alpha$, and Pa$\alpha$ broad lines \citep[e.g.,][]{ 2014MNRAS.445.3073P, 2018Natur.563..657G, Song2020,Song2021}.

Figures \ref{fig:f7} shows the periodic variation of BELs for the BBH-DB scenario with offset angles $\Delta i=0^\circ, 30^\circ, 60^\circ$, and $90^\circ$ in left to right columns, respectively. For an observer viewing the BBH systems with $i_{\rm BBH}=85^\circ$, the observed continuum variation is independent of the offset angle. While for the circumbinary BLR with an initial opening angle of $\theta_{\rm o}^{\rm ini}=30^\circ$, increasing $\Delta i$ corresponds to decreasing $i_{\rm BLR}$ and full width half maximum (FWHM) of BEL profiles, and the profile shapes changing from double-peaked/strong asymmetric ones to Lorentz-like single peaked ones. Correspondingly, the maximum amplitude of the DB enhancement/weakening happens to the coplanar case ($\Delta i=0^\circ$), with increasing offset angle, less BLR clouds are enhanced/weakened by the DB effect efficiently, and the minimum amplitude appear in the perpendicular case ($\Delta i=90^\circ$). 

As to the periodic variation of profile shapes, the edge-on viewed BLR present double-peaked or strongly asymmetric shapes with the blue and red parts raising up or going down alternatively. With increasing $\Delta i$, the BEL profiles are observed in more like asymmetric shapes with increasing $\Delta i$, and finally become Lorentz-like shapes at $\Delta i=90^\circ$ ($i_{\rm BLR}\sim 0^\circ$). When the BLR is observed at $i_{\rm BLR}\sim 30^\circ$ to $60^\circ$, whether the double-peaked/asymmetric features appear depend on the value of $\Delta i$, the ratio of $i_{\rm BLR}/a_{\rm BBH}$, and the initial opening angle of BLR $\theta_{\rm o}^{\rm ini}$. 

Different from the case that viewing a BLR at different $i_{\rm BLR}$ at the coplanar case as shown in Figure 8 of PAPER I, the variation of the $i_{\rm BLR}$ in this work is caused by the increasing offset angles, hence the close to face-on viewed BLR at $\Delta i=60^\circ$ and $90^\circ$ show broadened wings than that in Figure 8 of PAPER I due to the orbital precession. Unlike the periodically varying profile wings in the low $\Delta i$ cases (e.g. $0^\circ$ and $30^\circ$), the wings caused by the orbital precession have no significant flux variation as shown in the bottom row of Figure \ref{fig:f7}.

For the BBH-IntDB scenario shown in Figure \ref{fig:f8}, the BEL profile variation is modulated periodically in the time-delay direction (Figure \ref{fig:f6}), which cause a systematic enhancement/weakening of BEL fluxes, overlapped with periodic enhancement/weakening of the blue- and red-shifted parts of the profile resulted by the DB effect. The amplitudes of profile variation at different $i_{\rm BLR}$ are comparable to each other because of the isotropic radiation of the dominant intrinsic variation. 

Given the same periodic light curves of continuum radiation, the BBH mass in the BBH-IntDB case is systematically smaller than that in the BBH-DB case, hence the BEL profiles of the two scenarios have different characteristics in two aspects: 1) the FWHM of the BBH-DB model is systematically larger than that of the BBH-IntDB model, 2) the face-on viewed BLR present more like Gaussian shapes for the BBH-IntDB model, instead of the Lorentz shapes for the BBH-DB model, since the BBH-IntDB model with larger $R_{\rm BLR}/a_{\rm BBH}$ has less significant feature of broad wings caused by orbital precessions.

\subsection{BEL profile variation for a misaligned BLR with different compositions}

\begin{figure*}[ht]
\centering
\includegraphics[width=1.0\textwidth]{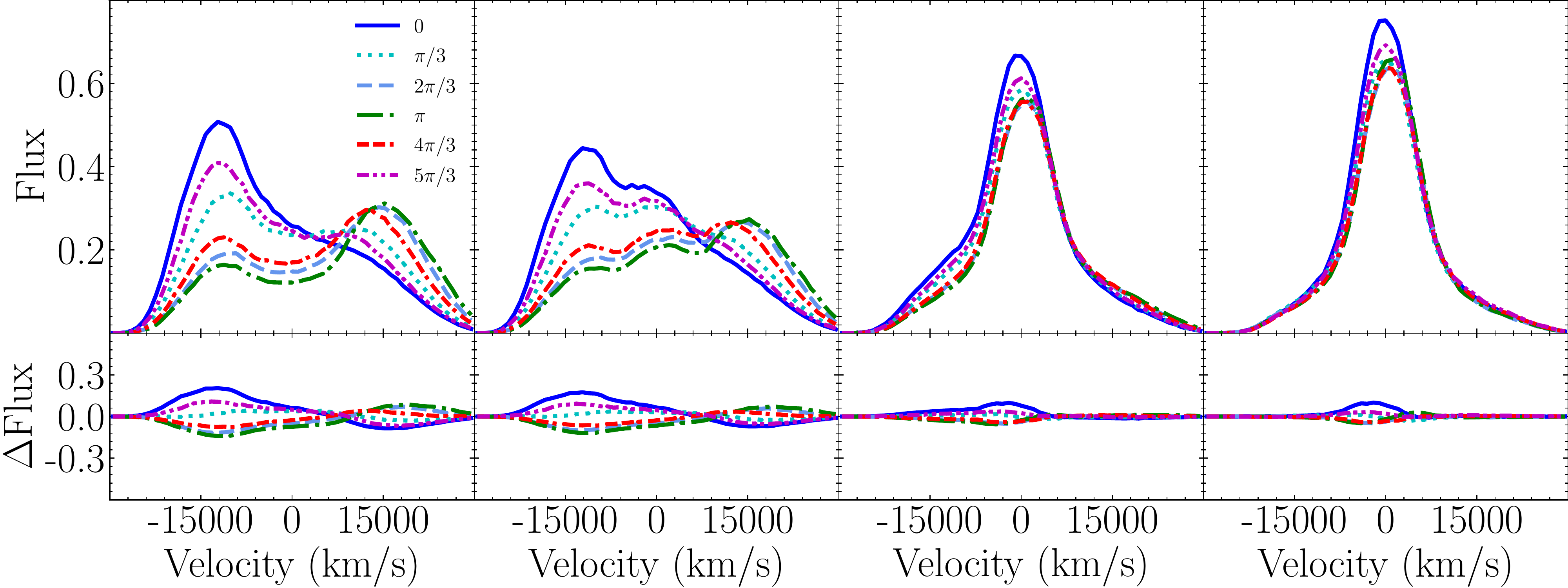}
\caption{Variations of BEL profiles over a single period (2 yr) of continuum variation for the BBH-DB scenario with four BLR configurations, i.e.,  $M_{\rm BLR,c}:M_{\rm BLR,v} = 1:0$ (left column), $5:1$ (middle-left column), $1:5$ (middle-right column), and $0:1$ (right column). In top panels we plot the profiles of the same six phases as shown in Figure \ref{fig:f7}, and the corresponding bottom panels show the difference of the line profile at each phase from the mean line profile.}
\label{fig:f9}
\end{figure*} 

\begin{figure*}[ht]
\centering
\includegraphics[width=1.0\textwidth]{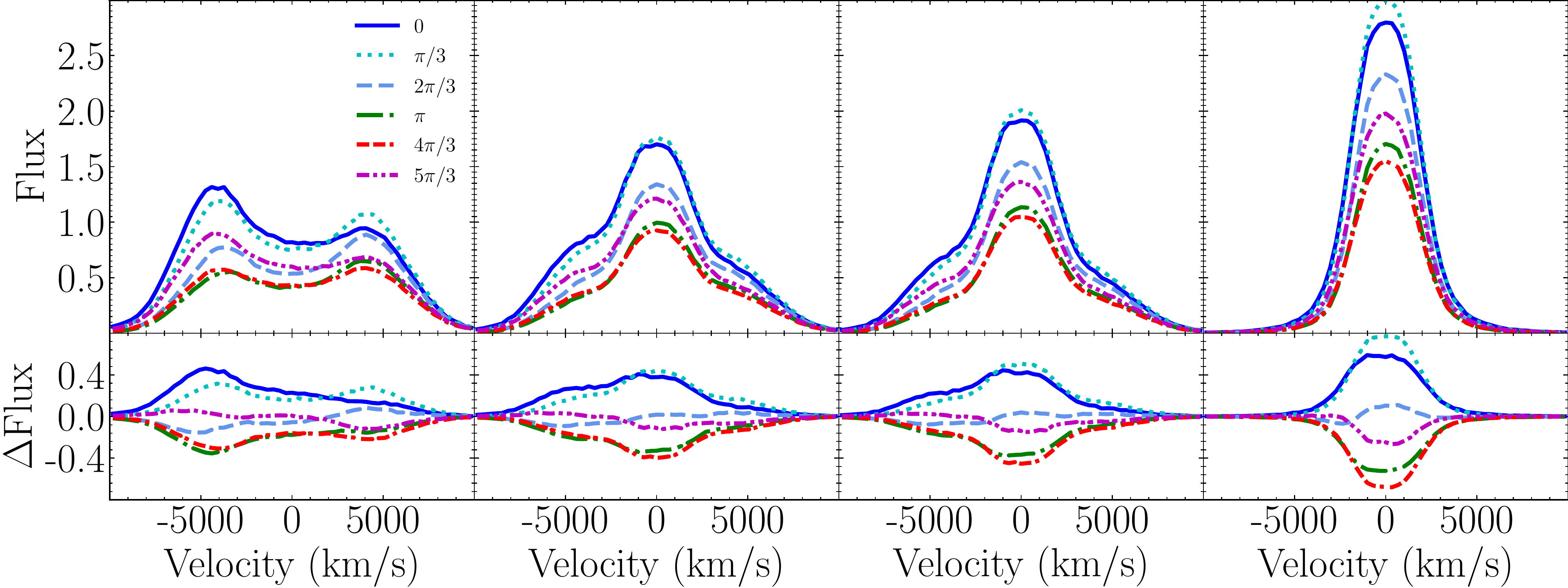}
\caption{
Variations of BEL profiles over a single period (2 yr) of continuum variation for the BBH-IntDB scenario with four BLR configurations, i.e.,  $M_{\rm BLR,c}:M_{\rm BLR,v} = 1:0$ (left column), $5:4$ (middle-left column), $4:5$ (middle-right column), and $0:1$ (right column). Legends are similar to Figure \ref{fig:f9}.
}
\label{fig:f10}
\end{figure*} 

In the above calculations and analyses, we assume that all BLR clouds distribute around a uniform BLR middle plane with an initial opening angle of $30^\circ$, however, the circumbinary BLR might consist of two components, which come from the two separate BLRs associate with each of the two BHs when they are on a much larger separation during the BBH evolution process \citep[e.g.,][]{ 2018ApJ...862..171W, 2020ApJS..247....3S, 2020A&A...635A...1K}. 

To investigate the feature of BELs emitted from a circumbinary BLR that consists of different components, and study how the BEL profiles vary with these different combinations, we assume four kinds of circumbinary BLRs for simplicity to explore how the BEL profiles vary with different BBH-BLR systems. By setting a BLR component with its middle plane aligned with the BBH orbital plane ($\rm BLR_c$ with its number of clouds $N_{\rm BLR,c}$), and another vertical component with its middle plane perpendicular to the BBH orbital plane ($\rm BLR_c$ with its number of clouds $N_{\rm BLR,v}$), the four BLR configurations are constructed as follows: 

\begin{itemize}
\item[(1)] a circumbinary BLR aligned with the BBH orbital plane, i.e., $N_{\rm BLR,c}:N_{\rm BLR,v} = 1:0$, which is the same as the $\Delta i=0^\circ$ case shown in Figures \ref{fig:f7} and \ref{fig:f8} (also our PAPER I). 
\item[(2)] the circumbinary BLR consists of both $\rm BLR_{\rm c}$ and $\rm BLR_{\rm v}$ components, with the ratio of cloud numbers follow the mass ratio of the two BHs, i.e., $N_{\rm BLR,c}:N_{\rm BLR,v} = 5:1$ and $5:4$ for the BBH-DB and   BBH-IntDB scenarios, respectively. 
\item[(3)] the circumbinary BLR have the two BLR components with $N_{\rm BLR,c}:N_{\rm BLR,v} = 1:5$ and $4:5$ for the BBH-DB and BBH-IntDB scenarios,  respectively.
\item[(4)] a circumbinary BLR perpendicular to the BBH orbital plane, i.e., $N_{\rm BLR,c}:N_{\rm BLR,v} = 0:1$,  the same as the right column of Figures \ref{fig:f7} and \ref{fig:f8} with $\Delta i=90^\circ$.
\end{itemize}
Both the $\rm BLR_c$ and $\rm BLR_v$ components are set to have the same initial opening angle $\theta_o^{\rm ini}=30^\circ$ as assumed in Table \ref{tbl:t1}.

Figure \ref{fig:f9} show the variation of BEL profiles emitted from the above four type BLR configurations for the BBH-DB scenario. The shape of a BEL profile is determined by the number fraction of the two perpendicular components $N_{\rm BLR,c}:N_{\rm BLR,v}$. Columns from left to right show an increasing number fraction of the vertical BLR component. For $N_{\rm BLR,c}:N_{\rm BLR,v}=1:0$ (left column), the double-peaked profiles at the six phases are the same as shown in the left panel of Figure \ref{fig:f7}. When the number fraction of the vertical BLR component increases to $N_{\rm BLR,c}:N_{\rm BLR,v}=5:1$ (middle-left panel), in which the coplanar BLR component still dominates the emission, a small bump at the line center appear with its flux variation following the double-peaked profile, since the DB effect takes the maximum effect to the coplanar case and minimum to the vertical case. Once the vertical BLR component dominating the total emission with $N_{\rm BLR,c}:N_{\rm BLR,v}=1:5$ (middle-right panel), the profile shape is Lorentz-like and similar to the one in the right panel ($N_{\rm BLR,c}:N_{\rm BLR,v}=0:1$), but has a bump in the blue wing contributed by the coplanar BLR component and hence has larger flux variation than the red wing.

For the BBH-IntDB scenario, which requires higher mass ratio of the BBH system than the BBH-DB model to produce similar optical/UV light curve of the continuum radiation, Figure \ref{fig:f10} shows quite different profile shapes and flux variations from those shown in Figure \ref{fig:f9}. Compared to the coplanar case ($N_{\rm BLR,c}:N_{\rm BLR,v}=1:0$, left column), BEL profiles resulted by $N_{\rm BLR,c}:N_{\rm BLR,v}=5:4$ present both features of the coplanar and vertical BLR components, with the blue and red wings close to the coplanar case, but the line center more close to the vertical case (right panel). Since the intrinsic variation caused time-delay effect dominates the enhanced/weakened emissivity of BLR clouds, adding the vertical BLR component makes both the blue wing and the line center having large flux fluctuation. With increasing number fraction of the vertical BLR component ($N_{\rm BLR,c}:N_{\rm BLR,v}=4:5$, middle-right panel), the Gaussian-like shaped component at the line center takes more significant role on shaping the profile and also amplitude of the flux variation, accompanying with the weakened signal and variation of blue and red wings.

\section{Discussions}

Previous studies have focused on the 2DTFs of BLR clouds and BEL profiles emitted from two separate BLRs surrounding each of the two BHs at larger separations, i.e., $R_{\rm BLR}/a_{\rm BBH}<1$ \citep[e.g.,][]{2010ApJ...725..249S, 2018ApJ...862..171W, 2020ApJS..247....3S, 2020A&A...635A...1K}. The configuration of a BLR-BBH system, especially for disk-like BLRs, can be structured by a series of parameter combinations, such as opening angles and inclination angles of the BLR, BLR size or bolometric luminosity of the two BHs, the kinematics of the BLR clouds, i.e., inflow, outflow, or Keplerian orbit. Different configurations of BLRs can actually result BEL profiles in complex shapes \citep[see a review by][]{2020RAA....20..160W}.

For the geometry of a circumbinary BLR outside the BBH system ($R_{\rm BLR}/a_{\rm BBH}>>1$), we have also discussed the possible variation caused by different setup of BBH orbital period ($T_{\rm orb}$, continuum variation period ($T_{\rm var}$), and spectral index $\alpha$ in the far-ultraviolet and optical bands (see PAPER I for details). As predicted by comparing the modelling of light curves and BEL profile fittings \citep[e.g.,][]{Song2020, Song2021}, the misaligned BLRs from BBH orbital plane make the emissivity of BLR clouds more complex than the coplanar case. 

With constructed BBH-DB and BBH-IntDB models by assuming the same $T_{\rm orb}$, $T_{\rm var}$, and $\alpha$ with BLR clouds rotating in elliptical orbits (see Table \ref{tbl:t1} for details), the emission of BLR clouds mainly vary with the following parameters:
\begin{itemize}
\item For a BBH system with smaller BLR size, e.g., $R_{\rm BLR}/a_{\rm BBH}\sim 4.6$ in the BBH-DB model (Figure \ref{fig:f7}), orbital precessions of BLR clouds can increase the opening angle of the BLR and thus the shape of BEL profiles. The significance of orbital precession becomes less important for larger $R_{\rm BLR}/a_{\rm BBH}$, e.g., the BBH-IntDB with $R_{\rm BLR}/a_{\rm BBH}\sim 9.8$.
\item The profile shape is not only affected by the offset angles and $i_{\rm BLR}$, but also by the initial BLR opening angle $\theta_{\rm o}^{\rm ini}$. Larger $\theta_{\rm o}^{\rm ini}$ can make the double-peaked features appeared in the $\Delta i=60^\circ$ and $i_{\rm BLR}=35^\circ$ becomes more asymmetric, while smaller $\theta_{\rm o}^{\rm ini}$ can make the asymmetric shapes becoming double-peaked ones.
\item The existence of two or more BLR components make the profile emission more complex. If the ratio of cloud numbers for the two BLR components follows the mass ratio of the two BHs, then higher mass ratios require higher spectral resolution and signal to noise ratio to distinguish them, which would be very important for decomposing and modelling the BLR-BBH geometry.
\end{itemize}

\section{Conclusions}

In this paper, we investigate the properties of broad emission lines (BELs) from supermassive binary black hole (BBH) systems and its variation, under the assumption that the associated broad line region (BLR) is circumbinary and misaligned from the BBH orbital plane. The continuum emission from the systems (or specifically the accretion associated with the BBH secondary component) have periodic variations, either due to the relativistic Doppler boosting (DB) effect or due to the combination of intrinsic accretion rate variation (dominant) and DB effect, i.e., the BBH-DB and BBH-IntDB scenarios, respectively. With similar optical/UV light curves of continuum radiation produced by these two different scenarios, the response of BELs to the continuum source is expected to be a significant indicator for identifying them. Our main results on the BEL emission and variation are summarized as follows.

With increasing offset angles from the coplanar case (BLR co-aligned with BBH orbital plane) to the vertical case (BLR perpendicular to BBH orbital plane),
e.g., at a fixed viewing angle of $i_{\rm BBH}=85^\circ$,  
for the BBH-DB scenario, BEL profiles vary from double-peaked/strongly asymmetric shapes in the coplanar case ($\Delta i=0^\circ$, $i_{\rm BBH}=85^\circ$) to Lorentz shapes in the vertical case ($\Delta i=90^\circ$), with enhanced blue and red wings due to the orbital precession of BLR clouds close to the two BHs. The opening angle of a BLR significantly offset from BBH orbital plane at its inner region may be substantially larger than that at the outer region, mainly because the corotating BBH system induces significant orbital precession of BLR clouds at the inner region that those at the outer region, especially for the cases with large offset angles. The DB effect induced periodic enhancement/weakening to the blue- and red-sides of those BELs are the strongest when the BLR is aligned with BBH orbital plane, but the  weakest when BLR is perpendicular to BBH orbital plane. For the BBH-IntDB case, the ratio of BLR size to BBH semimajor axis is usually larger than that for the BBH-DB case, thus the orbital precession caused by the BBH system takes a less important role than that in the BBH-DB case. The BEL profiles hence vary from double-peaked features in the coplanar case to Gaussian like shapes in the vertical case. The amplitude of profile variation for the BBH-IntDB scenario is nearly independent of the varying offset angles because of the isotropic radiation for the dominant intrinsic variation. 

The circumbinary BLR may be composed of two components offset from each other, which are natural results from the merger of two BH systems each with a BLR when the separation of the two BHs are large. For example, the circumbinary BLR may be the combination of a component aligned with and a component offset from BBH orbital plane with different BLR cloud numbers. In this case, the periodic variation of BEL profiles in both the BBH-DB and BBH-IntDB scenarios have unique features, which would be helpful for identifying BBH systems, reveal the associated BLR structures, and infer the formation history of BBH systems.

\section*{Acknowledgments}
This work is partly supported by the National Key R\&D Program of China (Grant Nos. 2020YFC2201400, 2020SKA0120102, and 2016YFA0400704), the National Natural Science Foundation of China (Grant Nos. 11690024, 11873056, and 11991052), the Strategic Priority Program of the Chinese Academy of Sciences (Grant No. XDB 23040100), and the Beijing Municipal Natural Science Foundation (Grant No. 1204038).


\end{document}